\documentclass[prl,aps,twocolumn,showpacs,floatfix]
{revtex4}
\usepackage[dvips]{graphicx}
\usepackage{amsmath}
\usepackage{amssymb}

\begin{document}

\title{Observation of Unusual Mass Transport in Solid hcp $^4$He}

\author{M. W. Ray  and
R. B. Hallock}
\affiliation{Laboratory for Low Temperature Physics, Department of Physics,\\
University of Massachusetts, Amherst, MA 01003}

\date{\today}

\begin{abstract}
Solid $^4$He has been created off the melting curve by growth at
nearly constant mass via the ``blocked capillary" technique and
growth from the $^4$He superfluid at constant temperature. The
experimental apparatus allows injection of $^4$He atoms from
superfluid directly into the solid. Evidence for the
superfluid-like transport of mass through a sample cell filled
with hcp solid $^4$He off the melting curve is found. This mass
flux depends on temperature and pressure.
\end{abstract}

\maketitle

Experiments by Kim and Chan\cite{Kim2004a, Kim2004b, Kim2005,
Kim2006}, who studied the behavior of a torsional oscillator
filled with hcp solid $^4$He, showed a clear reduction in the
period of the oscillator as a function of temperature at
temperatures below T $\approx$ 250 mK.  This observation was
interpreted as evidence for the presence of ``supersolid" behavior
in hcp solid $^4$He. Subsequent work in a number of laboratories
has confirmed the observation of a period shift,
 with the interpretation
of mass decoupling in most cases in the 0.05 - 1 percent range,
but with dramatically larger decoupling seen in quench-frozen
samples in small geometries\cite{Rittner2007}.  Aoki et
al.\cite{Aoki2007} observed sample history dependence under some
conditions. These observations and interpretations, among others,
have kindled considerable interest and debate concerning solid hcp
$^4$He.

Early measurements by Greywall\cite{Greywall1977}, showed no
evidence for mass flow in solid helium.  Work by the Beamish group
also showed no evidence for mass flow in two sets of experiments
involving Vycor\cite{Day2005} and narrow channels\cite{Day2006}.
Sasaki et al.\cite{Sasaki2006} attempted to cause flow through
solid helium on the melting curve, using a technique similar to
that used by Bonfait et al.\cite{Bonfait1989} (that showed no
flow). Initial interpretations suggested that flow might be taking
place through the solid\cite{Sasaki2006}, but subsequent
measurements have been interpreted to conclude that the flow was
instead likely carried by small liquid regions at the interface
between crystal faces and the surface of the sample
cell\cite{Sasaki2007}, which were shown to be present for helium
on the melting curve. Recent work by Day and Beamish\cite{Day2007}
showed that the shear modulus of hcp solid $^4$He increased at low
temperature and demonstrated a temperature and $^3$He impurity
dependence very similar to that shown by the torsional oscillator
results. The theoretical situation is also complex, with clear
analytic predictions that a supersolid cannot exist without
vacancies (or interstitials)\cite{Prokofev2005}, numerical
predictions that no vacancies exist in the ground state of hcp
solid $^4$He\cite{Boninsegni2006,Clark2006,Boninsegni2006a}, and
{\it ab initio} simulations that predict that in the presence of
disorder the solid can demonstrate
superflow\cite{Boninsegni2006,Pollet2007,Boninsegni2007} along
imperfections. But, there are alternate points of
view\cite{Anderson2007}.  There has been no clear experimental
evidence presented for the flow of atoms through solid hcp $^4$He.

We have created a new approach, related to our
``sandwich"\cite{Svistunov2006} design, with an important
modification. The motivation was to attempt to study hcp solid
$^4$He at pressures off the melting curve in a way that would
allow a chemical potential gradient to be applied across the
solid, but not by squeezing the hcp solid lattice directly.
Rather, the idea is to inject helium atoms into the solid from the
superfluid. To do this off the melting curve presents rather
substantial experimental problems due to the high thermal
conductivity of bulk superfluid helium. But, helium in the pores
of Vycor, or other small pore geometries, is known to freeze at
much higher pressures than does bulk
helium\cite{Beamish1983,Adams1987,Lie-zhao1986}. Thus, the
``sandwich" consists of solid helium held between two Vycor plugs,
each containing superfluid $^4$He.

The schematic design of our experiment is shown in figure 1. Three
fill lines lead to the copper cell; two from room temperature,
with no heat sink below 4K, enter via liquid reservoirs, R1, R2,
atop the Vycor (1 and 2) and a third (3) is heat sunk at 1K and
leads directly to the cell, bypassing the Vycor. The concept of
the measurement is straightforward: (a) Create a solid sample Shcp
and then (b) inject atoms into the solid Shcp by feeding atoms via
line 1 or 2. So, for example, we increase the pressure on line 1
or 2 and observe whether there is a change in the pressure on the
other line.  We also have capacitive pressure gauges on the sample
cell, C1 and C2, and can measure the pressure {\it in situ}.  To
conduct the experiment it is important that the helium in the
Vycor, the liquid reservoirs atop the Vycor, and the lines that
feed the Vycor contain $^4$He that does not solidify. This is
accomplished by imposing a temperature gradient between R and Shcp
across the Vycor, a gradient which would present insurmountable
difficulties if the Vycor were not present. While the heat
conducted down the Vycor rods in our current apparatus is larger
than we expected, and this presently limits our lowest achievable
temperature, we have none the less obtained interesting results.

\begin{figure}
\resizebox{3.2 in}{!}{\includegraphics{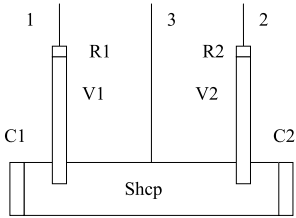}} \caption{Schematic
 of the apparatus.  Vycor rods, V1, V2, enter the
copper sample cell, with solid sample Shcp (4.45 cm long, 0.64 cm
dia.), which has capacitive pressure gauges, C1, C2 and is cooled
from the bottom. Thermometers are located on the cell and at the
reservoirs, R1, R2. The pressures in lines 1 and 2 are recorded at
room temperature. } \label{cell}
\end{figure}

To study the flow characteristics of our Vycor rods, we measured
the relaxation of pressure differences between line 1 and line 2
with superfluid $^4$He in the cell at $\sim$ 20 bar at 400 mK,
with the tops of the Vycor rods in the range 1.7 $<$ T$_1$ = T$_2$
$<$ 2.0 K, temperatures similar to some of our measurements at
higher pressures with solid helium in the sample cell. The
relaxation was linear in time as might be expected for flow
through a superleak at critical velocity. The pressure recorded by
the capacitive gauges shifted as it should. An offset in the
various pressure readings if T$_1$ $\ne$ T$_2$ was present due to
a predictable fountain effect across the two Vycor superleaks. Our
Vycor rods readily allow a flux of helium atoms, even for T$_1$,
T$_2$ as high as 2.8K.

To study solid helium, one approach is to grow from the superfluid
phase (using ultra-high purity helium, assumed to have $\sim$ 300
ppb $^3$He). With the cell at T $\approx$ 400 mK, we added helium
to lines 1 and 2 to increase the pressure from below the melting
curve to $\approx$ 26.8 bar. Sample A grew in a few hours and was
held stable for about a day before we attempted measurements on
it. Then the pressure to line 1, P1, was changed abruptly from
27.1 to 28.6 bar (figure 2). There resulted a gradual decrease in
the pressure in line 1 and a corresponding increase of the
pressure in line 2. Note that pressure can increase in line 2 only
if atoms move from line 1 to line 2, through the region of the
cell occupied by solid helium, Shcp. We also observed a change in
the pressure recorded on the capacitive pressure gauges on the
cell, e.g. C1 (C1 and C2 typically agree). As these pressure
changes evolved, we hoped to see the pressure in line 1 and line 2
converge, but the refrigerator stopped after 20 hours of operation
on this particular run.  Note that the change in P2 is rather
linear (0.017 bar/hr) and does not show the sort of non-linear
change with time that one would expect for the flow of a viscous
fluid. Our conclusion is that helium has moved through the region
of hcp solid $^4$He, while the solid was off the melting curve,
and that this flow from line 1 to line 2 was at a limiting
velocity, consistent with superflow. From the behavior of the
pressure gauges on the cell, it is clear that atoms were also
added to the solid.

\begin{figure}
\resizebox{3.2 in}{!}{\includegraphics{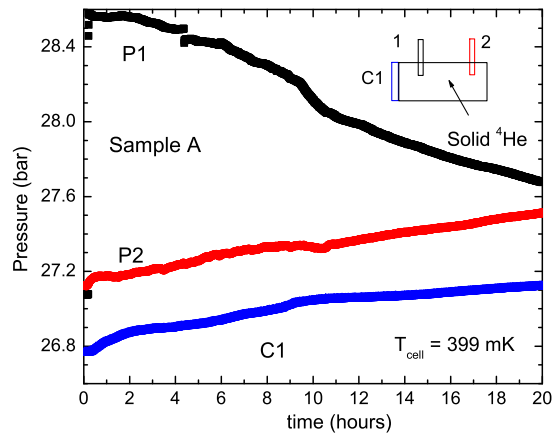}} \caption{(color
online) Response of the apparatus to a pressure step applied to
line 1 for sample A, grown at 26.8 bar. The shift in P1 just after
4 hours is an artifact from the fill of a nitrogen trap. The
regulator that slowly fed helium to line 1 was shut off at 6
hours.  The pressure increase in line 2 was essentially linear in
time, especially for the final ten hours.} \label{flow}
\end{figure}

We next grew a new solid sample, B, again by growth from the
superfluid, but we grew it at a faster rate and did not dwell for
a day prior to measurements. This sample also demonstrated flow,
with the pressure difference relaxing over about 5 hours after we
stopped adding atoms to line 1. The pressure step applied to line
1 was from 26.4 to 28.0 bar. While $^4$He was slowly added to line
1, P2 increased. After the addition of atoms was stopped, the
change in these pressures appeared to depend on P1-P2, with P2
showing curvature and regions of predominantly $\sim$ 0.076 and
$\sim$ 0.029 bar/hr. Next, we used the same solid sample and moved
it closer to the melting curve (1.25 K), but maintained it as a
solid, sample C. We applied a pressure difference by increasing
the pressure to line 1 from 26.0 to 28.4 bar, but in this case
there was no increase in P2; the pressure difference P1-P2
appeared nearly constant, with a slight increase in pressure
recorded in the cell. It is possible that this difference in
behavior is an annealing effect, but it may also be due to a
reduced ability to flow through the same number of conducting
pathways. Next, after a warm up to room temperature, we prepared
another sample, G, with P,T coordinates much like sample A, but
used a time for growth, and pause prior to injection of helium,
that was midway between those used for samples A and B. The
results again showed flow, (P2 changing $\sim$ 0.008 bar/hr; with
C1, C2 similar). Finally, we injected sample G again, but there
were some modest instabilities with our temperatures. A day later,
we injected again on this same sample, now two days old and termed
H; and then again, denoting it sample J. Short term changes were
observed in P1 and P2, but P1-P2 was essentially constant at
$\approx$ 1.38 bar for more than 15 hours. In another sequence, we
created sample M (like G), increased P1, observed flow, warmed it
to 800 mK, saw no flow, cooled it to 400 mK, increased P1, saw no
flow, decreased P1, saw flow, increased P1 again, and saw flow
again. (Typically if an increase in P1 shows flow, a decrease in
P1 will also show flow.) Yet another sample, Y, created similar to
A, showed linear flow like A, but when warmed to 800 mK showed no
flow. Whatever is responsible for the flow appears to change
somewhat with time, sample history, and is clearly dependent on
sample pressure and temperature.

How can we reconcile such behavior when the measurements of
Greywall and Day and Beamish saw no such flow\cite{Greywall1977,
Day2005, Day2006}? The actual explanation is not clear to us, but
there is a conceptual difference between the two types of
experiments: These previous experiments pushed on the solid helium
lattice; we inject atoms from the superfluid (which must have been
the case for the experiments of Sasaki et al.\cite{Sasaki2006}, on
the melting curve). If predictions of superflow along structures
in the solid\cite{Pollet2007, Boninsegni2007, Shevchenko1987}
(e.g. dislocations of various sorts or grain boundaries) are
correct, it is possible that by injecting atoms from the
superfluid we can access these defects at their ends in a way that
applying mechanical pressure to the lattice does not allow.

We have also grown samples via the ``blocked capillary" technique.
In this case the valves leading to lines 1 and 2 were controlled
and the helium in line 3 was frozen.
 Sample D was created this way and exited the melting curve in
the higher pressure region of the bcc phase and settled near 28.8
bar.  There then followed an injection of $^4$He atoms via line 1
(figure 3). Here we observed a lengthy period during which a
substantial pressure difference between lines 1 and 2 did not
relax, and to high accuracy we saw no change in the pressure of
the solid as measured directly in the cell with the capacitive
gauges C; C1 changed $<$ 0.0003 bar/hr. Behavior of this sort was
also observed for the same sample, but with a much smaller (0.21
bar) pressure shift, with no flow observed. And, warming this
sample to 900 mK produced no evidence for flow (sample E, not
shown). Four other samples (F, T, V, W) were grown using the
blocked capillary technique, with the lower pressure samples (T,
V) demonstrating flow. Pressure appears to be an important
variable, but not growth technique.

\begin{figure}
\resizebox{3.2 in}{!}{\includegraphics{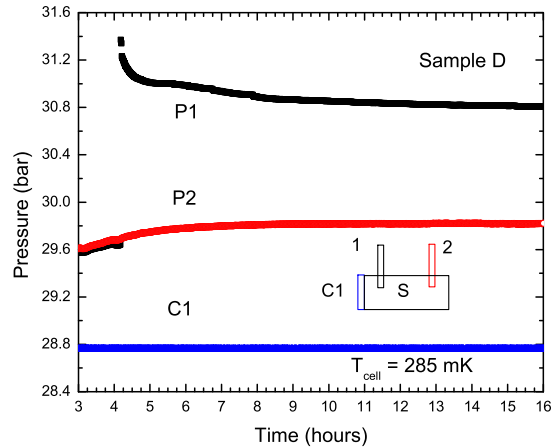}} \caption{(color
online) Behavior of a solid helium sample D grown with the blocked
capillary technique at 28.75 bar.  No flow is present after
several hours, with C1 unchanged.} \label{blocked}
\end{figure}

To summarize the focus of our work to date, on figure 4 we show
the location of some of the samples that we have created.  Samples
grown at higher pressure have not shown an ability to relax from
an applied pressure difference over intervals longer than 10 hr.;
they appear to be insulators. Samples grown at lower pressures
clearly show mass flux through the solid samples, and for some
samples this flux appears to be at constant velocity. Samples,
warmed close to 800 mK, and one warmed near to the melting curve
at 1.25 K, and a sample created from the superfluid at 800 mK all
showed no flow. We interpret the absence of flow for samples
warmed to or created at 800 mK to likely rule out liquid channels
as the conduction mechanism. Annealing my be present for the 1.25
K sample, but we doubt that this explains the 800 mK samples.
Instead we suspect that whatever conducts the flow (perhaps grain
boundaries or other defects) is temperature dependent. Sample
 pressure and temperature are important; sample history may be.

\begin{figure}
\resizebox{3.2 in}{!}{\includegraphics{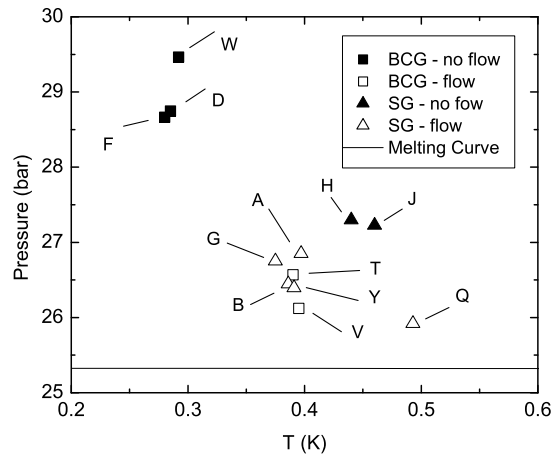}} \caption{Phase
diagram depicting some of the locations for which we have made
samples. BCG = blocked capillary growth; SG = growth from the
superfluid at constant temperature.  Solid symbols indicate
samples for which no flow was observed. Letters denote samples,
some of which are mentioned in the text.} \label{phase}
\end{figure}

The data of figure 2 can be used to deduce the mass flux through
and into the sample.  From that 20-hour data record we conclude
that over the course of the measurement 1 $\times$ 10$^{-4}$ grams
of $^4$He must have moved through the cell from line 1 to line 2,
and that about 4.5 $\times$ 10$^{-4}$ grams of $^4$He must have
joined the solid. If we write M/t = $\xi$$\rho$vxy, as the mass
flux from line 1 to line 2, where M is the mass that moved in time
t, $\rho$ is the density of helium, $\xi$ is the fraction of the
helium that can flow, v is the velocity of flow in the solid, and
xy is the cross section that supports that flow, we find $\xi$vxy
= 8 $\times$ 10$^{-9}$ cm$^3$/sec. We know from measurements on
the Vycor filled with superfluid that it should not limit the
flow. So, if we take the diameter of our sample cell (0.635 cm),
presuming that the full diameter conducts, we can deduce that
$\xi$v = 2.52 $\times$ 10$^{-8}$ cm/sec, which, if, for arbitrary
example, v = 100 $\mu$/sec, results in $\xi$ = 2.5 $\times$
10$^{-6}$.

An alternate approach is to presume instead that what is
conducting the flow from line 1 to line 2 is not the entire cross
section of the sample cell but rather a collection of discrete
structures (say, dislocation lines, or grain boundaries).  If this
were the case, with one dimension set at x = 0.5 nm, an atomic
thickness, then for the flow from line 1 to line 2, $\xi$vy = 0.16
cm$^2$/sec .  If we assume that $\xi$ = 1 for what moves along
these structures then vy = 0.16 cm$^2$/sec.  If we adopt the point
of view that what can flow in such a thin dimension is akin to a
helium film, we can take a critical velocity of something like 200
cm/sec\cite{Telschow1974}. In such a case, we find y = 8 $\times$
10$^{-4}$ cm. If our structures conduct along an axis, where the
axis is, say 0.5 nm x 0.5 nm, then we would need 1.6 $\times$
10$^4$ such structures to act as pipe-like conduits. This, given
the volume of our cell between our two Vycor rods (0.6 cm$^3$),
would require a density of such structures of at least 2.67
$\times$ 10$^4$ cm$^{-2}$, and roughly five times this number
(10$^5$ cm$^{-2}$) to carry the flux that also contributes mass to
the solid as its pressure increases.

We have conducted experiments that show the first evidence for
flow of helium through a region containing solid hcp $^4$He off
the melting curve. The phase diagram appears to have two regions.
Samples grown at lower pressures show flow, with flow apparently
dependent on sample history, with reduced flow for samples at
higher temperature, which is evidence for dependence on
temperature. Samples grown at higher pressures show no clear
evidence for any such flow for times longer than 10 hours. The
temperatures utilized for this work are well above the
temperatures at which much attention has been focused, but
interesting behavior is seen. Further measurements will be
required to establish in more detail how such behavior depends on
pressure and temperature, and on sample history, and the relevance
(if any) of our observations to the torsional oscillator and shear
modulus experiments that were conducted at lower temperatures.

We thank B. Svistunov and N. Prokofev for illuminating
discussions, which motivated us to design this experiment. We also
thank S. Balibar and J. Beamish for very helpful discussions and
advice on the growth of solid helium, M.C.W. Chan, R.A. Guyer, H.
Kojima, W.J. Mullin, J.D. Reppy, E. Rudavskii and Ye. Vekhov for
discussions. This work was supported by NSF DMR 06-50092, CDRF
2853, UMass RTF funds and facilities supported by the
NSF-supported MRSEC.



\end{document}